\documentclass{article} 
\usepackage{graphicx}
\usepackage{epstopdf}
\usepackage{dcolumn}
\usepackage{bm}
\usepackage{epsfig}
\usepackage{authblk}
\pdfoutput=1

\begin{document}
\title{Experimental characterization of electron beam driven wakefield modes in a dielectric woodpile Cartesian symmetric structure}
\author[1]{P.D.Hoang \thanks{hoangpd@g.ucla.edu}}
\author[1]{G.Andonian}
\author[1]{I.Gadjev}
\author[1]{B.Naranjo}
\author[1]{Y.Sakai}
\author[1]{N.Sudar}
\author[1]{O.Williams}
\affil[1]{Department of Physics and Astronomy, University of California, Los Angeles, California 90095-1547, USA}
\author[2]{M.Fedurin}
\author[2]{K.Kusche}
\author[2]{C.Swinson}
\affil[2]{Accelerator Test Facility, Brookhaven National Laboratory, Upton, NY 11973, USA}
\author[1]{J.B.Rosenzweig}
\affil[1]{Department of Physics and Astronomy, University of California, Los Angeles, California 90095-1547, USA}

\date{\today}

\begin{abstract}
The use of photonic structures in the terahertz (THz) spectral region may enable the essential characteristics of confinement, modal control, and electric field shielding for very high gradient accelerators based on wakefields in dielectrics. We report here an experimental investigation of THz wakefield modes in a 3D photonic woodpile structure. Selective control in exciting or suppressing of wakefield modes with non-zero transverse wave vector is demonstrated by using drive beams of varying transverse ellipticity.  Additionally, we show that the wakefield spectrum is insensitive to the offset position of the strongly elliptical beam. These results are consistent with analytic theory and 3D simulations, and illustrate a key advantage of wakefield systems with Cartesian symmetry, the suppression of transverse wakes by elliptical beams.
\end{abstract}
\maketitle
Dielectric wakefield accelerators (DWA) driven by short-pulse, intense charged particle beams have recently emerged in the THz frequency regime \cite{thompson2008}. In this burgeoning class of advanced accelerator, fields orders of magnitude beyond the current state of the art in radiofrequency (RF) linear accelerators – to greater than the GV/m level – are obtained. This progress is due in part to the availability of high-brightness, ultra-short pulse electron beams, and also to advances in few-micron resolution fabrication techniques. Recent experimental results on DWA using cylindrical dielectric lined waveguides have demonstrated sustained average accelerating gradients in excess of 1.3 GeV/m as reported in \cite{OShea2016}. Attaining fields in an accelerator based on solid state matter in this spectral range is a significant achievement for advanced acceleration techniques. However, many further aspects of the DWA need to be experimentally addressed to bring the viability of the DWA methods towards promising applications, e.g. in future linear colliders \cite{barish2008} and compact light sources \cite{emma2010}. 

Even before the onset of dielectric breakdown \cite{thompson2008}, the upper bound of useful acceleration gradient is limited by high field effects on the dielectric material \cite{OShea2016,dolgaleva2015}.  To minimize these deleterious effects the entry of strong electric field into the dielectric from both the intense beam's space-charge fields and the electromagnetic mode excited in the beam's wake must be suppressed. Further, aside from materials considerations, one typically seeks to increase the DWA  gradient by using a smaller aperture size $a$ following the  (Cherenkov) scaling $E_{z} \propto Q/a^2$  \cite{Tremaine1997,Barov2000}. However, transverse wakefields, which can cause spurious collective transverse motion that leads to beam breakup instabilities, scale as $E_{\perp} \propto Q/a^3$, and thus beam stability at THz frequencies presents a formidable challenge \cite{Li2014}. One promising solution for mitigating transverse wakefields uses structures that possess Cartesian instead of cylindrical symmetry, with one dimension (x) having variations that are small on the scale of the mode wavelength. In the 2D case where there is no x-dependence in both the structure and the beam, the resultant wakefields have, in the ultra-relativsitic limit, zero net (i.e, magnetic and electric) transverse force \cite{Tremaine1997}, as a direct consequence of the assumed symmetry. Some recent DWA studies on slab-symmetric metal-clad dielectric structures \cite{Andonian2012,Antipov2012} have successfully demonstrated accelerating mode excitation and concomitant acceleration of charge.

While much progress has been made in demonstrating high gradient, efficient acceleration in DWAs, many issues remain unaddressed experimentally. One must, for example, develop structures that partially suppress the penetration of electric fields into the dielectric, as is critically important in very high gradient applications when materials encounter nonlinearities and breakdown limits \cite{OShea2016,dolgaleva2015}. This may be accomplished by two key design features. First, one may do away with the flat longitudinal vacuum/dielectric interface present in the simple slab case, which enforces continuity of $E_z$, and achieve a field reduction in the dielectric by a factor as large as $1/\epsilon$ through the implementation of the boundary conditions. Further, photonic structures can be used to confine modes within the accelerating vacuum channel. An example was demonstrated in \cite{Andonian2014}  where a pair of 1D photonic Bragg reflectors were used to confine the modes formed between the slabs. Use of higher dimensional photonic structures having more sophisticated shapes can offer yet more capabilities for tuning mode properties \cite{Naranjo2012}, and controlling the variation of the field in all directions, potentially placing spurious modes outside of the band gap, permitting them to propagate away from the beam \cite{Smirnova2005a}, and yielding a mode that is optimized for acceleration applications.  

\begin{figure}	
\includegraphics[width=1\columnwidth]{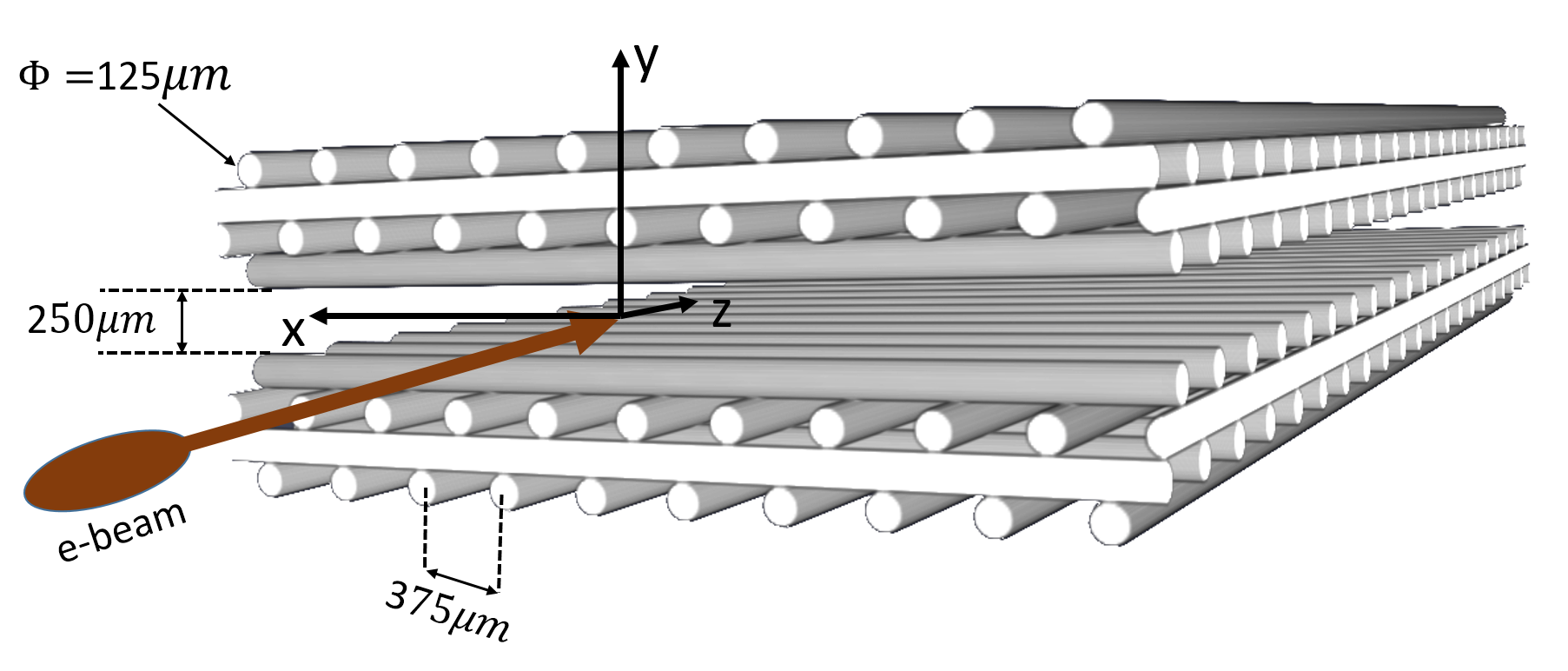}
\caption{\label{fig:fig1} Computer rendered model of the woodpile structure with detailed dimension.}
\end{figure}
\begin{figure}
\includegraphics[width=1\columnwidth]{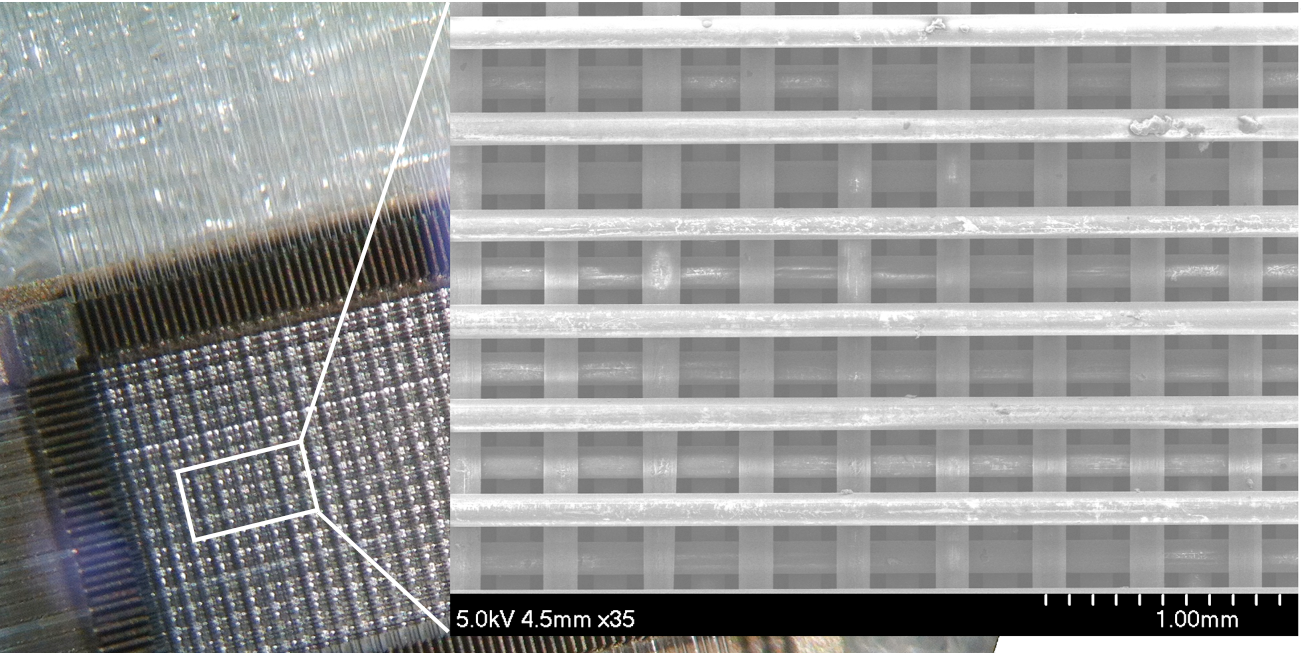}
\caption{\label{fig:fig2}Optical image showing sapphire rods in slotted holder during the manual fabrication process. Inlet shows an SEM image (top down view) of a section of the device.}
\end{figure}

In this Letter, we therefore present a wakefield study of a structure having Cartesian symmetries, constituted of discrete, periodic dielectric rods that are formed into a so-called woodpile geometry, schematically shown in Fig. \ref{fig:fig1}. The introduction of the woodpile lattice into this symmetry class marks, the first time that a 3D photonic crystal is incorporated into a DWA structure. In embarking on a comprehensive investigation of 3D photonic structures, the woodpile lattice, which entails relatively simple and well-known photonic properties, may serve as the first stepping stone from which systematic refinements can be made. The single-period woodpile structure studied here contains all basic modal ingredients from which more sophisticated structures can be understood. Our goal here is thus to experimentally demonstrate control over excitation of wakefield modes.

As a first experimental exploration of the wakefield structure excited in a 3D photonic environment, the woodpile was constructed without the features that allow for a full photonic band gap \cite{Cowan2008,Ho1994}. As is discussed below, wakefield modes are resonantly formed on the edge of the light cone; the modes extend into the surrounding vacuum due to the lack of strong field confinement and evanescent fields are present. This "leaky" confinement, in fact, works to our experimental advantage here. Modes, once formed, propagate down the length of the structure and are easily extracted into free space, due to diminished impedance mismatch. The coherent THz radiation that forms the wakefield mode is thus launched straightforwardly into a quasi-optical transport line leading to detection systems that measured its spectral properties.

The features of the structure can be seen in Fig. \ref{fig:fig1}. Each sapphire (Al$_2$O$_3$) rod is $125\mu m$ in diameter. Sapphire has long been a candidate for sub-mm DWA applications due to its mechanical robustness, high damage threshold \cite{England2014}, low loss, and a relatively high dielectric constant in the THz frequency range \cite{Grischkowsky1990a},  $\epsilon \approx 10.5$. Each of the two identical halves that makes up the structure consists of a full single woodpile stacking period \cite{Ho1994,Joannopoulos2008}. As seen in Fig. \ref{fig:fig1}, the main periodicity in both the wide transverse x- and the longitudinal z-direction is $375 \mu m$, and the vacuum gap that the beam traverses has a vertical clearance of 250 $\mu m$. The rods in the two rafts that are immediately below and above the drive beam run perpendicular to the beam axis, an orientation that serves to suppress electric field penetration into the dielectric while augmenting the longitudinal component of field in the defect region. There are 28 of these periods in the z-direction, equivalent to a length of 10.5 mm, and 16 in the x-direction, equivalent to 6 mm, in total. The structure was fabricated layer-by-layer (Fig. \ref{fig:fig2}) using a micro-rig holder with precision machined slots to facilitate the intricate arrangement. 

Unlike the simple slab structure whose normal modes (LSE/LSM) can be solved for analytically \cite{collin1990,mihalcea2012,xiao2001,jing2003}, the photonic nature of the woodpile requires numerical methods to obtain its modal structure $\omega(\vec{k})$. Wakefield modes are then solved for by invoking the Cherenkov condition \cite{Kremers2009}, $\omega(\vec{k}_\perp, k_z) = c k_{z}$, where the electrons are relativistic $|\vec{v}|=c$, and for clarity, the wave vector has been separated into its longitudinal and transverse components. While a point-like beam can excite all possible modes, we can limit coupling to large wave vectors by tailoring the Fourier content of the beam. Adjusting the beam bunch length, $\sigma_z$, allows us to select the highest $k_z$. Similarly, the beam width, $\sigma_x$, directly affects the contribution of $k_x$ to the wakefield spectrum \cite{Baturin2013}. The advantage of having large $\sigma_x$ is not only found in decoupling from unwanted non-zero $\vec{k}_\perp$ modes, but more importantly, also decoupling from net transverse forces, $\vec{F}_\perp \propto 1/\sigma^2_x$, as discussed for Cartesian symmetry structures in Ref. \cite{Tremaine1997}.

\begin{figure}
\includegraphics[width=1\columnwidth]{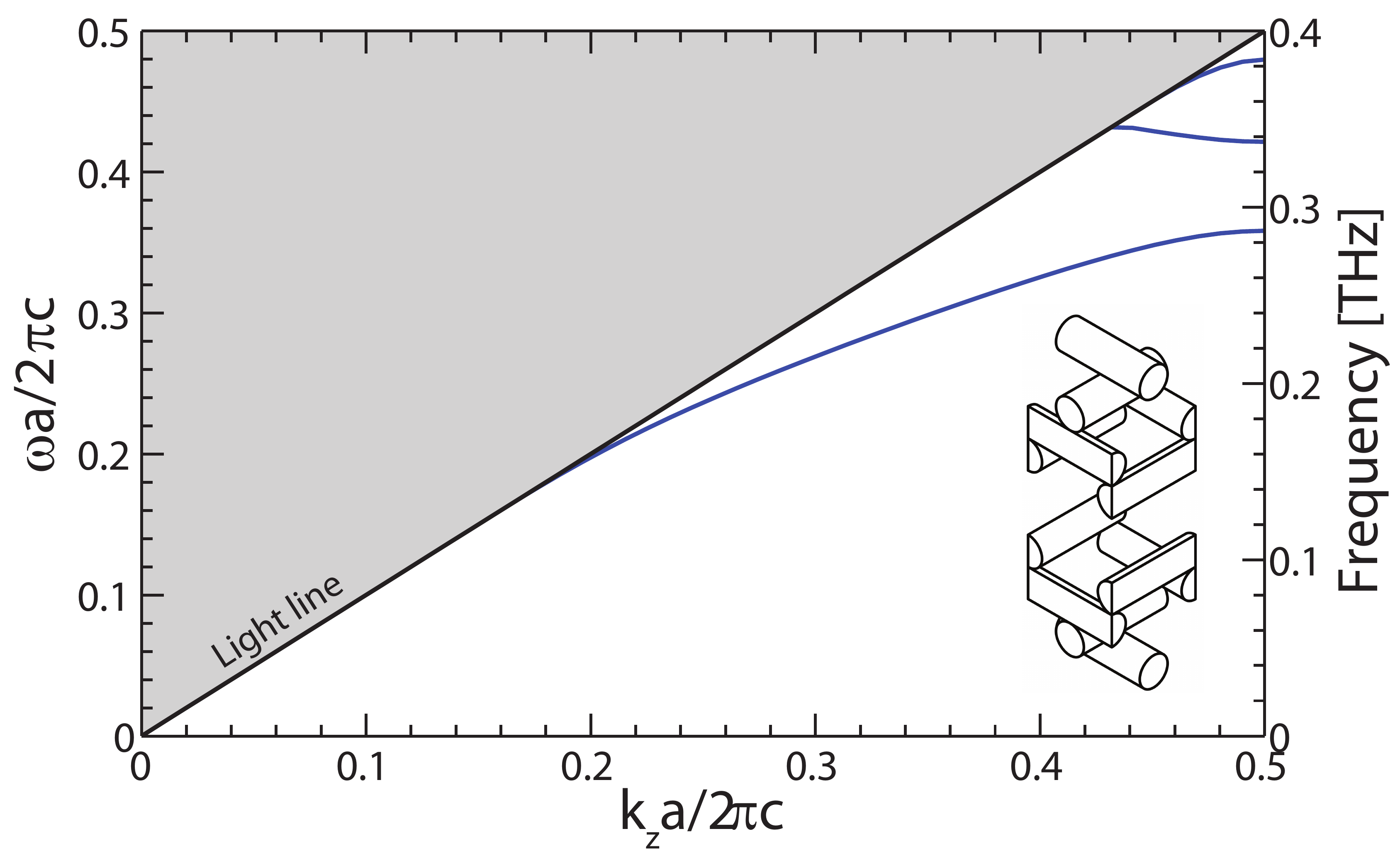}
\caption{\label{fig:fig3}Woodpile structure's dispersion shows three symmetric accelerating bands as a function of normalized longitudinal ($z$) wave vector. Wakefield modes are formed where the light line meets the bands, $\omega(k_z) = ck_z$. Left (right) vertical axis shows normalized (real) frequency; $a = 375 \mu m$. The inset shows the unit cell used for the eigenmode calculation.}
\end{figure}
\begin{figure}
\includegraphics[width=1\columnwidth]{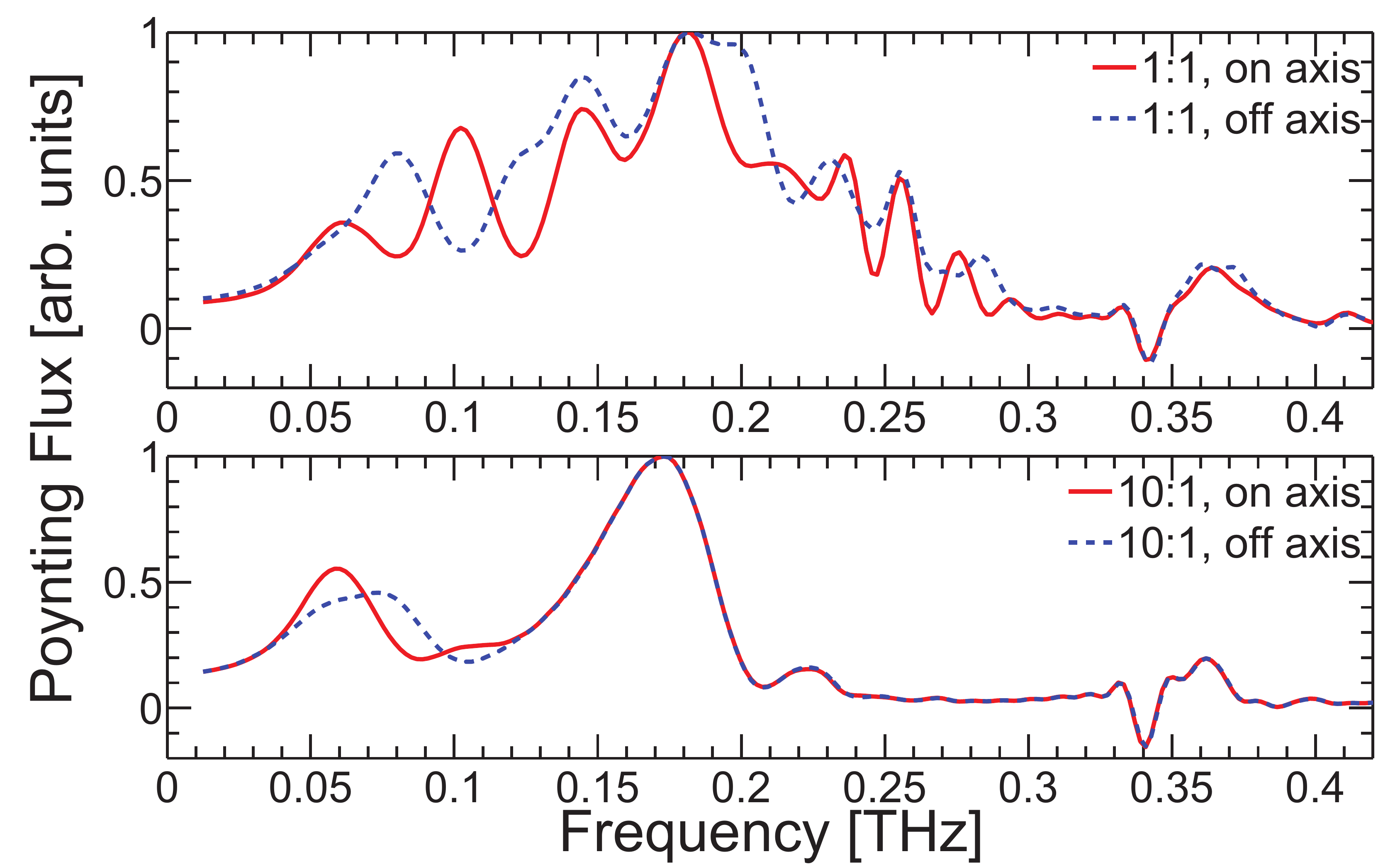}
\caption{\label{fig:fig4}(Color online) FDTD simulations showing spectral changes as a result of varying the beam's $\sigma_x/\sigma_y$ (1:1 and 10:1) and centroid offset (on- and off-axis) as labeled.}
\end{figure}
As the major emphasis of the presented experiments on the woodpile is an examination of the wakefield frequency spectrum, to orient the reader to the salient aspects of this photonic structure, we first introduce a computational analysis of the modes that can be excited via the wakefield mechanism. We begin by examining first the structure's frequency domain properties, and review the computational investigation of the most important structure eigenmodes. To perform this analysis \cite{Johnson2001}, shown in Fig. \ref{fig:fig3}, we make the following simplifications. First, we assume that the eigenmodes of the real finite structure are similar to those of an infinite (in x-z plane) but otherwise equivalent structure. This assumption affords use of the simple unit cell shown in the inset of Fig. \ref{fig:fig3}, thus avoiding the complications related to the use of a full super cell. Second, to emphasize first the dominant accelerating modes, particularly those preferentially excited by strongly elliptical $\sigma_x/\sigma_y \gg 1$ beams, we restrict our attention only to the eigenmodes with purely longitudinal wave vector, $\vec{k} = k\hat{z}$. Third, we further restrict the eigenmodes to only even symmetry (accelerating modes) defined as follows. The woodpile possesses two mirror symmetry planes ($x=0$ and $y=0$) where $\epsilon(\vec{r})=\epsilon(\mathcal{\bm{M}}\cdot\vec{r})$, and the mirror operator acting on a vector is defined as $\mathcal{\bm{M}}_{x}\cdot\langle x,y,z \rangle=\langle -x,y,z \rangle$ and $\mathcal{\bm{M}}_{y}\cdot\langle x,y,z \rangle =\langle x,-y,z \rangle$. For modes propagating in the z-direction, the field is either even $(+)$ or odd $(-)$, $\mathcal{\bm{M}}_{i}\cdot\vec{E}(\mathcal{\bm{M}}_{i}\cdot\vec{r})=\pm \vec{E}(\vec{r})$ where $i=x,y$. A symmetric mode is even under both mirror reflections. As we shall see momentarily, these assumptions are justified for drive beams with large aspect ratio. 

To demonstrate all these effects, and their manifestation as detectable changes to the wakefield spectrum, we have performed finite-difference time-domain (FDTD) simulations \cite{Oskooi2010}. The structures in these simulations are similar to the real device, and the beams were modeled as a relativistic $(|\vec{v}|=c)$ rigid current source with Gaussian distribution in all three dimensions. To obtain the wakefield spectrum, we calculated the Poynting flux through a plane that bisects the structure and is oriented normally to the direction of the beam. The area of the plane was made sufficiently large to capture all significant contributions to the electromagnetic waves in the structure.

In Fig. \ref{fig:fig4}, we summarized the wakefield Poynting flux for four different sets of beam parameters which are permutations of two aspect ratios, $\sigma_x/\sigma_y=1:1$ and 10:1, and two offset states, on- and off-axis, where the axis is defined by the symmetry line $x=y=0$, and with all parameters set to reflect the experimental scenarios discussed below. We see clearly that the wakefield spectrum excited by the 1:1 beam (top panel) is rich for both on- and off-axis driving beams. In contrast, the spectrum driven by 10:1 beam (bottom panel) is much simpler and is rather insensitive to beam displacement compared to the round beam case. This reflects the theoretical result that a large $\sigma_x$ (on the scale of the vacuum wavelength) serves to reduce the transverse wakefield \cite{Tremaine1997}.

Furthermore, with a few exceptions, the spectrum for the 10:1 cases resemble the predictions made in the band diagram for the symmetric (accelerating) modes. We can easily identify all three symmetric modes of Fig. \ref{fig:fig3} in the wakefield spectrum of Fig. \ref{fig:fig4}, \textit{i.e.} the fundamental at 185 GHz, the negative group velocity mode which shows up as negative Poynting flux at 350 GHz, and immediately followed by the third mode. Pertaining the same 10:1 cases, the finiteness of the structure and the beam in the time domain causes extraneous transient electromagnetic excitations unaccounted for in the frequency domain. Now referring back to the wakefield spectrum of the 1:1 on-axis case, we can still recognize the same three symmetric modes among the many that are excited. Since the on-axis beam cannot couple to transverse (asymmetric) modes, the additional modes observed in the time-domain must be wakefields having non-zero $\vec{k}_\perp$. The same can be stated about the 1:1 off-axis case, except that its spectrum also contains transverse wakefield modes. It is clear from this simulation exercise that the elliptical beam decouples from $\vec{k}_\perp$ modes and transverse deflection modes. 
\begin{figure}
\includegraphics[width=1\columnwidth]{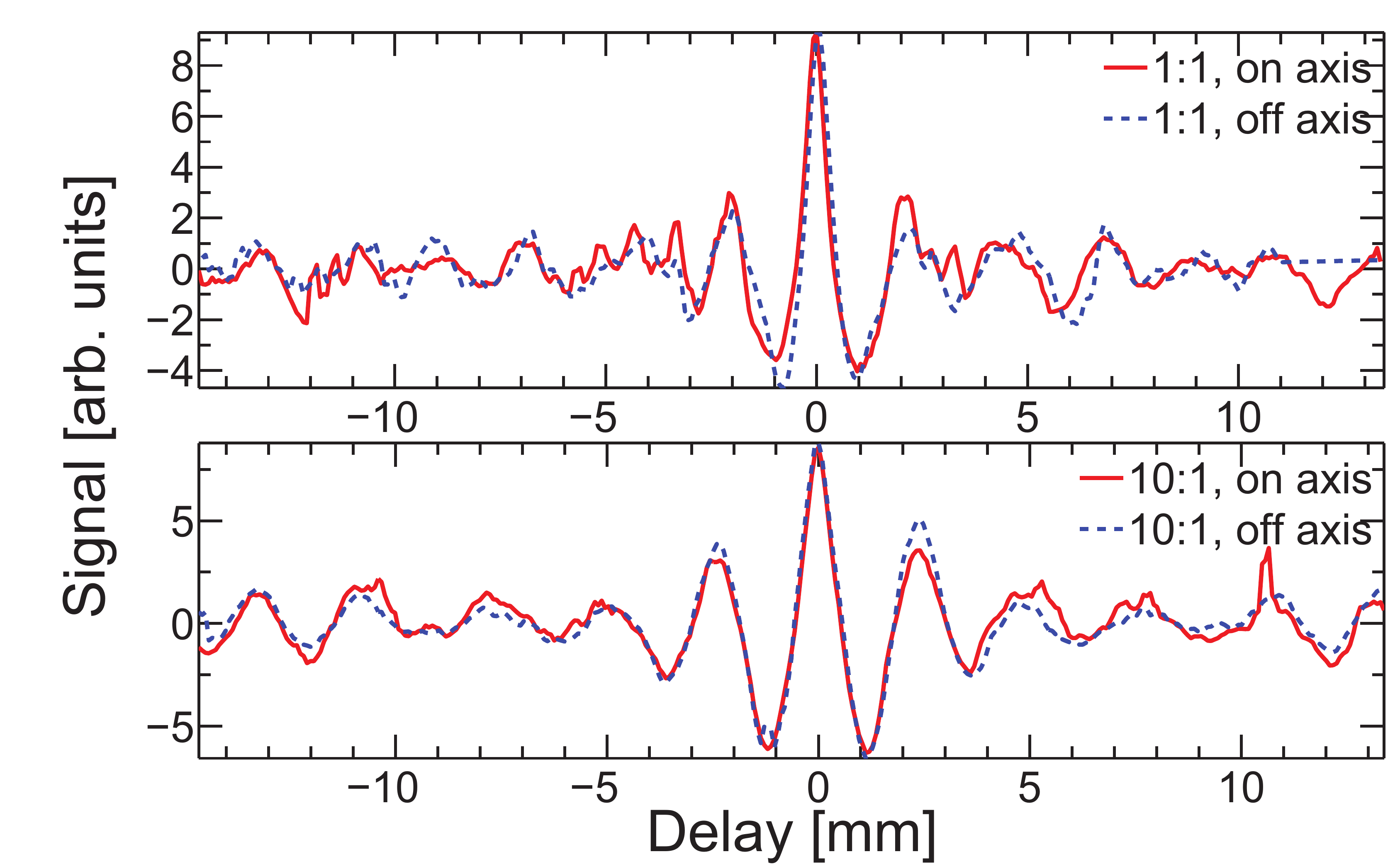}
\caption{\label{fig:fig5} (Color online) Field autocorrelation of the CCR signals measured by using Michelson interferometry for four sets of beam parameters as labeled.}
\end{figure}
\begin{figure}
\includegraphics[width=1\columnwidth]{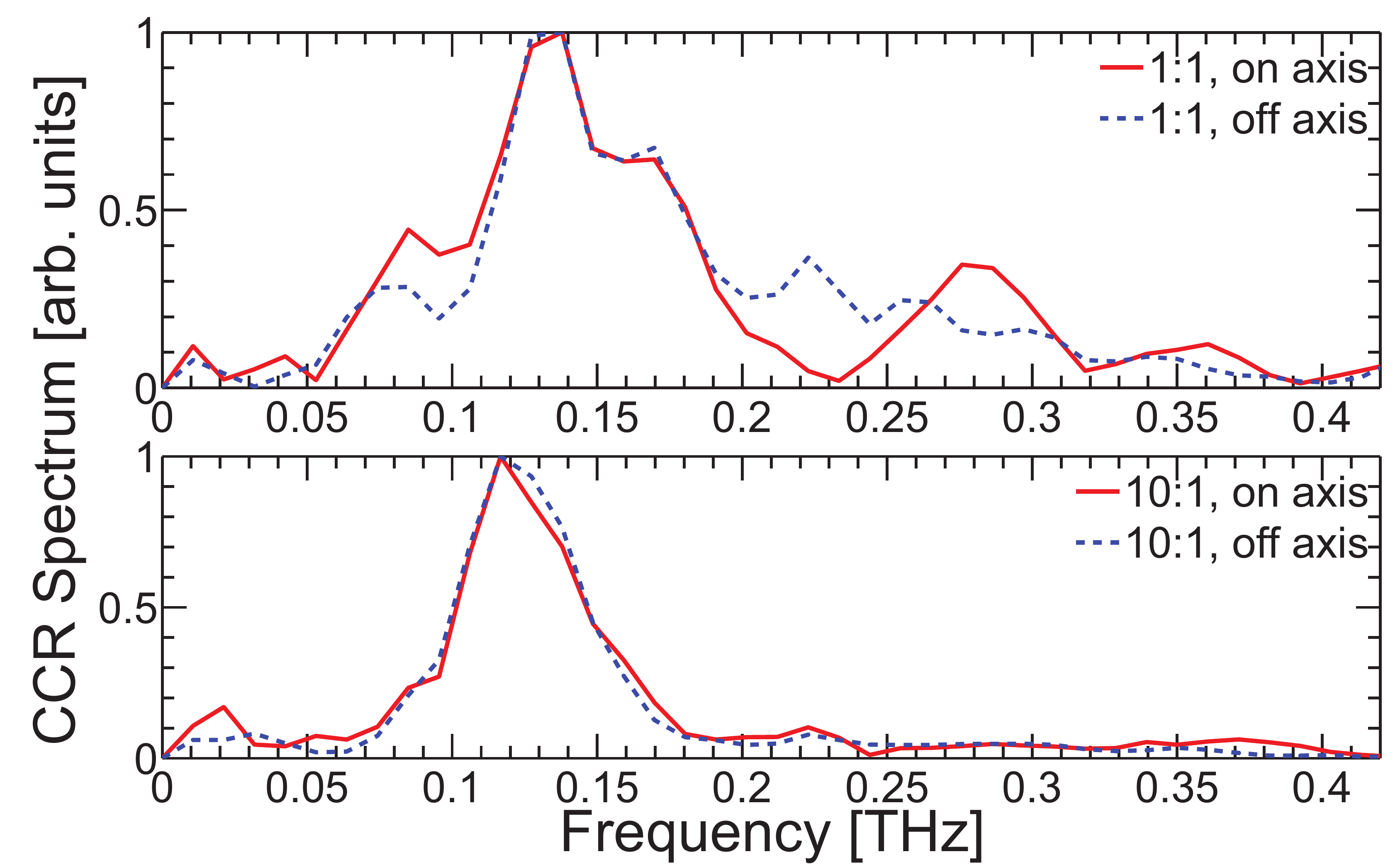}
\caption{\label{fig:fig6}(Color online). Measured CCR spectrum as directly obtained from Fourier transforming the signals in Fig. \ref{fig:fig5}.}
\end{figure}
The execution of the experiment generally followed the scheme outlined in the time-domain simulations. First, the wakefield spectrum was measured for the cases when the beams passed through the structure on-axis, and then the measurements were repeated with the beams displaced off-axis to observe the changes to the wakefield due to this asymmetric excitation. Considering the very tight vertical clearance of the vacuum gap (250$\mu m$), we were limited mostly to introducing a horizontal offset. In the end, a 1 mm horizontal offset was used. This was far enough from the axis to register substantial changes to the wakefield spectrum, and yet the beam was still far enough away from the edge of the structure to avoid direct impact. From $x-y$ coupling in the motorized stage that supports the structure, the 1 mm horizontal translation also served to introduce a 50$\mu m$ vertical offset. To measure the coherent Cherenkov radiation (CCR), which is the radiative form of the wakefield, we adopted the technique we have developed in previous DWA experiments performed \cite{Cook2009, Andonian2012, Andonian2014}. Measurements are made in the far field, and although we lose the symmetry information of the modes in the transport and detection scheme used, the frequency information is well preserved and can be easily measured using Michelson interferometry.

At the Brookhaven National Laboratory Accelerator Test Facility (BNL ATF), where the experiment was carried out, 57.6 MeV electrons bunches having normalized emittance $\epsilon_n=$2mm-mrad were focused to desired transverse beam sizes and waist location by a set of upstream quadrupole magnets. The beam aspect ratio, $\sigma_x/\sigma_y$, was measured by fitting its size on four beam profile monitors in the drift section downstream of the final magnetic focusing element. For the results presented in this article, two beam transverse ratios were established (corresponding to the FDTD simulations), 1:1 and 10:1. We refer to these cases as round and elliptical beams, respectively. To ensure passage of the beam through the structure, the beam vertical size at the waist was set to $\sigma_y \approx 50 \mu m$. Minimal beam loss due to scraping the structure was confirmed by a Faraday cup located downstream of the structure. It was determined that $\sim 10\%$ of charge was lost to the structure for both cases, and that on average the round and elliptical beam contained $150 pC$ and $235 pC$ of charge respectively. We note here that the difference in charge does not affect our results, as charge only contributes to the magnitude of the field and not its frequency content.

The frequency content is indeed set by the bunch length, which must be of sub-picosecond scale to ensure the coherent excitation of THz modes. We employed coherent diffraction radiation (CDR) \cite{Shibata1994} to non-destructively characterize the beam bunch length. A 2.5 mm aperture was used to generate CDR whose duration was measured using the Michelson interferometer. We determined by direct fitting \cite{Murokh1998452} that $\sigma_z \approx 250 \mu m$. This dimension was checked, and its invariance determined, regardless of aspect ratio. This guarantees that the beams always access the same spectral region, and that any differences in the wakefield are only due to the transverse dimensions and offsets introduced. In Fig. \ref{fig:fig5}, we show the four CCR field autocorrelation interferograms. As before, we categorized them into two groups according to aspect ratio. Periodicity observed on the wings of the interferograms indicates the presence of modes. We see clearly that the recordings for the 1:1 beam contain more oscillations and are more sensitive to offset than those for the 10:1 beam. In fact, this is even more obvious when we examine the obtained frequency spectrum of Fig. \ref{fig:fig6}. Again, the behavior is similar to what is observed in the Poynting flux in Fig. \ref{fig:fig4}. The 10:1 beam decouples from $\vec{k}_\perp$ modes as well as any transverse wakefield modes excited as a result of displacement. The absolute value of the frequency of the fundamental mode in the data while a bit lower than expected, falls within uncertainty of the determination of the permittivity of sapphire, as well as expected errors in the manual fabrication process. We also note that unlike the Poynting flux in simulation which captures all energy flow in both forward and backward directions, we only detect in the forward direction and hence cannot observe a backward propagating wave.

In summary, we have investigated properties of THz wakefields in a Cartesian dielectric woodpile, as a first exploration of the possibilities opened by 3D photonic structures in this emerging class of advanced, high gradient accelerator. We have observed in the wakefield spectrum while changing the beam transverse shape and position, and demonstrated control of mode excitation through changes in the beam aspect ratio. We found that a high aspect ratio beam preferentially couples to purely longitudinal wave vector modes. Furthermore, the wakefield spectrum driven by such beam was robust to displacement in stark contrast to the spectrum driven by a similar beam but with round aspect ratio, consistent with the expected advantageous property that the transverse deflecting wakefield is mitigated by use of a high aspect ratio beam. Both results confirm theoretical predictions and are corroborated by numerical simulations. They thus pave the way to use of more sophisticated structures--\textit{i.e.} of photonic type--along with high beam ellipticity in DWAs. Building upon this work, more preparation is under way for an experimental investigation of a full-fledged 3D structure with a complete band gap for high gradient field confinement. Such experiments, which will also critically permit exploration of suppression of field penetration into the dielectric are planned when the FACET II \cite{phinney2016} facility is available.

This work supported by the US DOE contract DE-SC0009914, and US Dept. of Homeland Security Grant 2014-DN-077-ARI084-01. Student stipend was partially provided by Office of Science Graduate Student Research (SCGSR) program. This research used resources of the Brookhaven National Laboratory Accelerator Test Facility, which is a DOE Office of Science user facility.

\bibliographystyle{plain}
\bibliography{woodpile_ref}

\end{document}